\documentclass[prd,superscriptaddress,a4paper,showpacs,showkeys,11pt,nofootinbib]{revtex4}
\usepackage{graphicx} % Required for inserting images
\usepackage{graphicx}
\usepackage{amsmath}% Required for inserting images
\usepackage{booktabs}
\usepackage{siunitx}
\usepackage{url}
\usepackage{hyperref}
\newcommand{\rr}{\mbox{\boldmath $r$}}
\newcommand{\rp}{\mbox{\boldmath $r^{\prime}$}}

\newcommand{\pv}{\mbox{\boldmath $p$}}

\newcommand{\kv}{\mbox{\boldmath $k$}}

\newcommand{\sv}{\mbox{\boldmath $s$}}

\newcommand{\Deltav}{\mbox{\boldmath $\Delta$}}

\newcommand{\dd}{\, \mathrm{d}}
\newcommand{\rb}{\mbox{\boldmath $b$}}

\begin{document}
\title{Investigating the inclusive $D^0$ photoproduction in ultraperipheral $PbPb$ collisions at the Large Hadron Collider}

\author{Victor P. {\sc Gon\c{c}alves}}
\email{barros@ufpel.edu.br}
\affiliation{Institute of Physics and Mathematics, Federal University of Pelotas (UFPel), \\
  Postal Code 354,  96010-900, Pelotas, RS, Brazil}

\author{Luana {\sc Santana}}
\email{luanas1899@gmail.com}
\affiliation{Institute of Physics and Mathematics, Federal University of Pelotas (UFPel), \\
  Postal Code 354,  96010-900, Pelotas, RS, Brazil}

\author{Wolfgang  {\sc Sch\"afer}}
\email{Wolfgang.Schafer@ifj.edu.pl}
\affiliation{The Henryk Niewodnicza{\'n}ski Institute of Nuclear Physics %(IFJ)
\\ Polish Academy of Sciences (PAN), 31-342 Krak\'ow, Poland
}

\begin{abstract}
The inclusive $D^0$ photoproduction in $PbPb$ collisions at the center - of - mass energies of the Large Hadron Collider (LHC) is investigated considering the color dipole $S$ - matrix approach. The analytical expressions for the differential distributions are derived in the impact parameter and transverse momentum spaces and predictions for the rapidity and transverse momentum distributions are presented considering three distinct models for the unintegrated gluon distribution of the nuclear target. In particular, we compare the predictions derived assuming a linear dynamics, with and without the inclusion of nuclear effects, with those obtained by solving the running coupling Balitsky - Kovchegov equation. A comparison of these predictions with the recent (preliminary) CMS data is also performed. Our results indicate that a detailed analysis of this observable will be very useful to improve our understanding of the strong interaction theory at high energies and in a nuclear medium.
\end{abstract}


\maketitle

\section{Introduction}
The study of the nuclear structure and the dynamics of the strong interactions in ultraperipheral heavy - ion collisions (UPHICs) became a reality in the last decade (for reviews see \cite{upc}). In particular, exclusive processes (e.g. the exclusive vector-meson photoproduction), where both incident particles remain intact,  were largely investigated over the last years, with a reasonable improvement of the theoretical description and the releasing of large amounts of data from distinct experimental collaborations at RHIC and LHC. Such studies were mainly motivated by the possibility of constraining the description of QCD dynamics at high energies \cite{klein,gluon,Frankfurt:2001db}, and improving our understanding of the 3D quantum imaging of partons inside protons and nuclei.  In contrast, the experimental analysis of inclusive processes, where one of the incident hadrons is broken up, is still in the embryonal phase, the first data having been published only recently \cite{ATLAS:2024mvt}. Theoretically, the 
investigation of inclusive processes provide additional constraints on the description of the QCD dynamics and nuclear effects (See e.g. Refs. \cite{Klein:2000dk,Klein:2002wm,Goncalves:2003is,Goncalves:2004dn,Goncalves:2009ey,Goncalves:2013oga,Goncalves:2015cik,Kotko:2017oxg,Goncalves:2017zdx,Guzey:2019kik,Guzey:2018dlm,Eskola:2024fhf,Goncalves:2019owz}), which are complementary to those derived in exclusive processes.

Motivated by the recent preliminary CMS Collaboration data \cite{CMSpreliminary} and the ongoing analysis in the  ALICE Collaboration \cite{ALICE}, we will perform in this paper the investigation of the inclusive $D^0$ meson photoproduction in ultraperipheral $PbPb$ collisions at the LHC (For a related study see Ref. \cite{Gimeno-Estivill:2025rbw}). We will derive the associated differential cross - section considering the color dipole $S$ - matrix dipole approach, proposed many years ago in Refs.  \cite{Nikolaev:2003zf,Nikolaev:2004cu,Nikolaev:2005dd,Nikolaev:2005zj,Nikolaev:2005qs}
   and recently extended for the gauge boson production in hadronic collisions \cite{Bandeira:2024zjl,Bandeira:2024jjl}. We will demonstrate that this approach reproduces, in the impact parameter space, the expression used in Ref. \cite{Goncalves:2017zdx} to derive, for  the first time, the predictions for the inclusive heavy meson photoproduction in UPHICs. Moreover, we will present the expression for the differential cross - section in the momentum space, where the behavior of the distribution is driven by the unintegrated gluon distribution (UGD) of the nuclear target, which is sensitive to treatment of the QCD dynamics and nuclear effects (For recent reviews about these subjects see, e.g. Refs. \cite{hdqcd,Klasen:2023uqj}). In our analysis, we will consider two models for the UGD based on a linear dynamics \cite{Jung:2004gs,Blanco:2019qbm}, with and without the inclusion of nuclear effects, and will compare the predictions with those derived in Refs. \cite{ALbacete:2010ad,Albacete:2012xq} using the solution of the running coupling  Balitsky - Kovchegov (rcBK) equation \cite{Balitsky:1995ub,Kovchegov:1999yj} for a nuclear target, which takes into account of nonlinear effects in the QCD dynamics. Such a comparison will allow us investigate the sensitivity of the observable on the description of the nucleus. A comparison with the preliminary CMS data will also be performed. As we will demonstrate, our results indicate that a future precise experimental analysis of the inclusive $D^0$ photoproduction in UPHICs will be very useful to improve our understanding of the QCD dynamics and about the description of the nuclear effects.

This paper is organized as follows. In the next Section, we present a brief review of the formalism needed to describe the  inclusive $D^0$ photoproduction in UPHICs. In particular, we will derive the differential distribution using the color dipole $S$- matrix approach and present the corresponding expressions in the impact parameter and momentum spaces. In Section \ref{sec:results} we will present our predictions for the rapidity and transverse momentum distributions, derived considering three distinct models for the nuclear UGD. Moreover, a comparison with the preliminary CMS data will be performed. Finally, in Section \ref{sec:summary} we will summarize our main conclusions.

\section{Formalism}

\subsection{$D$ - meson production in ultraperipheral heavy-ion collisions}
In what follows, we will consider ultraperipheral heavy ion collision, characterized by an impact parameter larger than the sum of the radius of incident ions ($b > R_A + R_B$). As we will focus on the inclusive photon - hadron production, where one of the incident ions breakup due to the strong interaction, in what follows we will assume that one of the ions as the photon source (ion $A$) and the other as the target (ion $B$). Moreover, we will assume that the nucleus that emits the photon is moving in direction of positive
rapidities. 

The  differential cross - section for the production of $D^0$ meson with transverse momentum $p_{T,D}$ at rapidity $Y_D$ can be factorized in terms of the cross-section for the charm production convoluted with the corresponding fragmentation function $D_{c/D^0}$, as follows
\begin{eqnarray}
    \frac{d^2 \sigma(A + B \rightarrow A \otimes D^0 + X)}{dY_D d^2p_{T,D}} = \int_{z_{\min}}^1 \frac{dz}{z^2} D_{c/D^0}(z, \mu^2) \left[ \frac{d^2 \sigma(A + B \rightarrow A \otimes c\bar{c} + X)}{dY_c d^2p_{T}} \right]_{p_{T} = \frac{p_{T,D}}{z}}
    \label{Eq:spectrum}
\end{eqnarray}
where  $\otimes$ represents the presence of a rapidity gap in the final state, associated with the photon exchange, and
%\begin{eqnarray}
%  \frac{d \sigma [h_1 h_2 \to h_1 + Q \bar Q +X]}{dx_Q d^2 p_T} &=& \int  dx dz_Q\, \delta(x_Q - x z_Q) f^{\rm eff}_{\gamma/A}(x) \frac{d \sigma (\gamma h_2 \to Q \bar Q X)}{d z_Q d^2 p_T}  \nonumber \\
%  &=& \int_{x_Q}^1 \frac{dz_Q}{z_Q} f^{\rm eff}_{\gamma/A}\Big( \frac{x_Q}{z_Q}\Big) \frac{d \sigma (\gamma h_2 \to Q \bar Q X)}{d z_Q d^2 p_T} \nonumber \\
%  &=& \int_{x_Q}^1 \frac{dx}{x} f^{\rm eff}_{\gamma/A}(x) \frac{d \sigma (\gamma h_2 \to Q \bar Q X)}{d z_Q d^2 %p_T} \Big|_{z_Q = \frac{x_Q}{x}}
%\end{eqnarray}
%We can multiply by $x_Q$:
\begin{eqnarray}
   \frac{d \sigma [A B \to A \otimes c \bar c +X]}{dY_c d^2 p_T} &=& \int_{x_c}^1 dx dz_c \, \delta(x_c - x z_c) xf^{\rm eff}_{\gamma/A}(x) z_c\frac{d \sigma (\gamma B \to c \bar c X)}{d z_c d^2 p_T} \nonumber \\ 
  &=& \int_{x_c}^1 dz_c \frac{x_c}{z_c} f^{\rm eff}_{\gamma/A}\Big(\frac{x_c}{z_c}\Big) \frac{d \sigma (\gamma B \to c \bar c X)}{d z_c d^2 p_T} \,\,,
  \end{eqnarray}
 with 
 \begin{eqnarray}
     x_c = \sqrt{\frac{p_T^2 + m_c^2}{s_{NN}}}\exp(+Y_c)\,\,.
 \end{eqnarray}
 Here $m_c$ is the charm mass and $\sqrt{s_{NN}}$ is center - of - mass of the nuclear collision.  Moreover, the effective nuclear photon flux, $f^{\rm eff}_{\gamma/A}$, is given by 
 %{\bf there was an error in eqn below, which dimensions did not add up.(jacobian missing) VP: I agree!}
 \begin{eqnarray}
x f^{\rm eff}_{\gamma/A}(x) = %\int_{b>2R_{Pb}}
 \int d^2\rb \, \omega N_{A}(\omega,\rb)P(\rb) P_{\rm strong}(\rb)    
 \end{eqnarray}
where $x = 2 \omega/\sqrt{s_{NN}}$, with $\omega$ being the photon energy and $P_{\rm strong}(\rb), P(\rb)$ represent ``survival probabilities'' against strong interactions and electromagnetic dissociation
respectively. In particular $P_{\rm strong}(\rb)$ excludes configurations where nuclei come into contact, and is approximated by us as $  P_{\rm strong}(\rb) = \theta(b - 2 R_{Pb}), b = |\rb|$.
%%%
%\begin{equation}
%    P_{\rm strong}(\rb) = \theta(b - 2 R_{Pb}) \, . 
%\end{equation}
%%%
The factor $P(\rb)$ ensures that no electromagnetic dissociation of the photon emitting ion occurs. We give its explicit form in Section \ref{sec:results}.
As impact parameters $b<2 R_{Pb}$ are cut off by the survival probablilty $P_{\rm strong}(b)$ for the purposes of this work it is sufficient  to use the photon spectrum for the pointlike source, 
which reads \cite{upc}
%{\bf VP A square was missing in the last two terms} \cite{upc}
%$N(\omega,b)$, can be expressed in terms  in terms of the charge form factor $F(q)$ as follows \cite{upc}
\begin{eqnarray}
    \omega N_A(\omega,\rb) = \frac{Z^2 \alpha_{\rm em}}{\pi^2} \, \frac{1}{b^2} \, \, \, \left( \frac{\omega b}{\gamma} \right)^2 K_1^2 \Big(\frac{\omega b}{\gamma}\Big) \, . 
\end{eqnarray}
%%%
%\begin{eqnarray}
% N_A(\omega,b) = \frac{Z^{2}\alpha}{\pi^2}\frac{1}{b^{2} v^{2}\omega}
%\cdot \left[
%\int u^{2} J_{1}(u) 
%F\left(
% \sqrt{\frac{\left( \frac{b\omega}{\gamma_L}\right)^{2} + u^{2}}{b^{2}}}
% \right )
%\frac{1}{\left(\frac{b\omega}{\gamma_L}\right)^{2} + u^{2}} \mbox{d}u
%\right]^{2} \,\,,
%\label{Eq:fluxo0}
%\end{eqnarray}
where $\alpha_{\rm em}$ is the electromagnetic fine structure constant, $\gamma$ is the Lorentz factor, which here is to be taken as $\gamma = \gamma_{\rm cm} \equiv \sqrt{s_{NN}}/(2m_N)$. 
%$v$ is the nucleus velocity. 
%In what follows, we will estimate the photon flux using a pointlike form factor. One has that for $P(\rb) = 1$, it will be given by
Putting $P(\rb)=1$, one obtains the well-known result
\cite{upc} 
%{\bf VP checked!}:
\begin{equation}
    \label{fluxonucleo}
    f^{\rm eff}_{\gamma/A}(x)= \frac{2Z^2\alpha_{\rm em}}{\pi x}\left [\xi K_0(\xi)K_1(\xi)-\frac{\xi^2}{2}\left ( K_1^2(\xi)-K_0^2(\xi) \right )  \right ],
\end{equation}
where $\xi=2R_{Pb}xm_N$, { with the nuclear radius being given by $R_A=1.2A^{1/3}-0.86A^{-1/3}$ (in fm).}
%and $\gamma_L$ is the Lorentz factor.
%{\bf value of $R_{Pb}$ used in calculations?}
The main input in the calculation of the $D$ - meson differential distributions in UPHICs, described by  Eq. (\ref{Eq:spectrum}), is the modeling of the charm photoproduction cross - section, which will be described in the next subsection using the color dipole $S$ - matrix approach.

\subsection{Heavy quark photoproduction cross-section in the color dipole $S$ - matrix approach}
In this subsection we will present a brief review of the color dipole $S$ - matrix approach, proposed in Refs. \cite{Nikolaev:2003zf,Nikolaev:2004cu,Nikolaev:2005dd,Nikolaev:2005zj,Nikolaev:2005qs}, and extended for the gauge boson production in hadronic collisions in Refs. \cite{Bandeira:2024zjl,Bandeira:2024jjl}. One has that the heavy quark photoproduction at high energies is  determined by the $\gamma g \rightarrow c \bar{c}$ subprocesses. In the laboratory frame, such a process  can be viewed  as an excitation of the perturbative $|c\bar{c} \rangle$ Fock state of the physical projectile $|\gamma\rangle$ by a one - gluon exchange with the target proton \cite{Nikolaev:1994de,Nikolaev:1995ty}. At high energies, the photon can be assumed to propagate along a straight line with a fixed impact parameter. The perturbative transition $\gamma \rightarrow c\bar{c}$ can be described in terms of the Fock state expansion for the physical state $|\gamma\rangle_{phys}$, which at the lowest order is given by \cite{Nikolaev:2003zf}
\begin{eqnarray}
|\gamma\rangle_{phys} = |\gamma\rangle_0 + \Psi(z,\rr)|c\bar{c}\rangle_0 \,\,,
\end{eqnarray} 
where $|...\rangle_0$ refers to a bare photon and $\Psi(z_c,\rr)$ is the probability amplitude to find the $c\bar{c}$ system with separation $\rr$ in the two-dimensional impact parameter space. Moreover, with $z$ the fraction of the longitudinal momentum of photon carried by the charm.  The  impact parameter of the photon is assumed to be $\rb$, which implies $\rb_c = \rb + z\rr$ and $\rb_{\bar{c}} = \rb - z\rr$.
One has that the corresponding cross-section for the charm production is a particular case of the master formula derived in Refs. \cite{Nikolaev:2003zf,Nikolaev:2004cu} for the $a \rightarrow bc$ transition, and is given by
\begin{eqnarray}
\label{Eq:Master_dijet}
\frac{d\sigma_{T,L} (\gamma \rightarrow c(p_c) \bar{c}(p_{\bar{c}}))}{dz d^2\kv d^2\Deltav} & = & \frac{1}{2 (2\pi)^4}\, 
\int  d^2\rr d^2\rr'  \exp[ i \kv \cdot (\rr - \rr^{\prime}) ] \Psi_{T,L}(z,\rr) \Psi^{*}_{T,L}(z,\rr') \nonumber \\
&\times&\int d^2\sv \exp[i \Deltav \cdot \sv] %\nonumber \\
%&\times&
 \Big\{
 \sigma(\sv+z\rr') + \sigma(\sv-z\rr) 
- \sigma(\sv -z(\rr-\rr') ) - \sigma(\sv)
\Big\}  \,\,, 
\end{eqnarray}
where $T (L)$ corresponds to a photon with transverse (longitudinal) polarization, $\kv = (1-z) \pv_c - z \pv_{\bar{c}}$ is the light-cone relative momentum, which is the conjugate variable to $\rr - \rr'$, and  $\Deltav = \pv_c + \pv_{\bar{c}}$ is the transverse momentum decorrelation. Moreover,  $\rr$ and $\rp$ are the $c\bar{c}$ transverse separations in the amplitude and its conjugated, respectively. Finally,
$\sigma$ is the dipole cross - section \cite{Nikolaev:1990ja}, which describes the $c\bar{c}$ - target  interaction and is determined by the QCD dynamics.

In this paper, we will focus on the inclusive charm production,  which has a cross - section that can be derived  integrating the above equation over the transverse momentum of the anticharm in the final state or, equivalently,  integrating  Eq. (\ref{Eq:Master_dijet})  over $\Deltav$. As a consequence, the corresponding cross - section will be given by
\begin{eqnarray}
\label{Eq:isolated}
\frac{d\sigma_{T,L} (\gamma h \rightarrow c\bar{c}X)}{dz d^2\pv} & = & \frac{1}{(2\pi)^2}\, 
\int d^2\rr \dd^2\rp \exp[i\pv \cdot (\rr - \rp)]\, \Psi_{T,L}(z,\rr) 
\Psi^{*}_{T,L}(z,\rp) \nonumber \\
 & \times & \frac{1}{2}\left[ \sigma(\rr,x) + \sigma(\rp,x) - 
 \sigma(|\rr - \rp|,x)\right]\,, 
\end{eqnarray}
with  $\pv = \pv_c$ and $x$ is the Bjorken variable. The products of wave functions are given by
\begin{eqnarray}
    \Psi_{T}(z,\rr) \Psi^{*}_{T}(z,\rp) & = & \frac{N_c\alpha_{em}e_c^2}{2\pi^2}\left\{\left[z^2+(1-z)^2\right]\epsilon ^2K_1(\epsilon r)K_1(\epsilon r')\frac{\mathbf{r}\cdot \mathbf{r}'}{rr'} + m_c^2K_0(\epsilon r)K_0(\epsilon r')\right\}
     \nonumber \\
    \Psi_{L}(z,\rr) \Psi^{*}_L(z,\rp)  & = & \frac{N_c\alpha_{em}e_c^2}{2\pi^2}\left\{ 4z^2(1-z)^2Q^2K_0(\epsilon r)K_0(\epsilon r')\right\} \,\,,  
\end{eqnarray}
where $Q^2$ is the photon virtuality, $\epsilon^2 = m_c^2 + z(1-z)Q^2$ and the $K_i$ functions are Bessel modified functions.

In our case, we are interested in the photoproduction limit, where the longitudinal contribution vanishes. Using the above expression for the tranverse photon wave functions, the charm  momentum spectrum can be expressed as follows
\begin{align*}
    \frac{\mathrm{d}\sigma_T(\gamma h\rightarrow c\bar{c}X
    )}{dz d^2\pv}=\frac{N_c\alpha_{em}e_c^2}{2\pi^2}&\left\{\left[z^2+(1-z)^2\right]\epsilon^2\mathcal{A}_1(z,p_T,\epsilon)+\left[zm_c+(1-z)m_c\right]^2\mathcal{A}_2(z,p_T,\epsilon)\right\}
\end{align*}
where
\begin{align*}
    \mathcal{A}_1(z,p_T,\epsilon)=&\frac{p_T}{\epsilon(p^2_T+\epsilon^2)}I_3(z,p_T)-\frac{1}{2\epsilon^2}I_1(z,p_T)+\frac{1}{4\epsilon}I_2(z,p_T)\\
    \mathcal{A}_2(z,p_T,\epsilon)=&\frac{1}{p^2_T+\epsilon^2}I_1(z,p_T)-\frac{1}{4\epsilon}I_2(z,p_T)
\end{align*}
with the auxiliary functions $I_i$ being defined by 
\begin{eqnarray}
I_1(z,p_T) & = & \int r\mathrm{d}r\;J_0(p_Tr)K_0(\epsilon r)\sigma(\mathbf{r},x)\\
        I_2(z,p_T) & = & \int \mathrm{d}r\;r^2J_0(p_Tr)K_1(\epsilon r)\sigma(\boldsymbol{r},x)\\
        I_3(z,p_T) & = & \int \mathrm{d}r\; rJ_1(p_Tr)K_1(\epsilon r)\sigma(\mathbf{r},x)    
\end{eqnarray}
Such expression for the transverse momentum spectrum was used in Ref. \cite{Goncalves:2017zdx} to derive the first predictions for the photoproduction of $D^0$ mesons in UPHICs. In this paper, we will consider an equivalent representarion of the spectrum, derived in the momentum space, which allow us to estimate the cross - section in terms of the unintegrated gluon distribution. Using the relation between the dipole-proton cross-section and unintegrated gluon distribution $\mathcal{F}(x,\mathbf{k})$ given by
\begin{align}
    \sigma(\mathbf{r},x)=&\frac{4\pi}{3}\int \frac{\mathrm{d}^2\mathbf{k}}{k^2}\alpha_s\mathcal{F}(x,\mathbf{k})\left(1-e^{i\mathbf{k}\cdot \mathbf{r}}\right)=\int \mathrm{d}^2\mathbf{k}f(x,\mathbf{k})\left(1-e^{i\mathbf{k}\cdot \mathbf{r}}\right)
\end{align}
where $f(x,\mathbf{k})=({4\pi\alpha_s}/{3k^2})\mathcal{F}(x,\mathbf{k})$ results that the spectrum will be given by
\begin{align*}
    \frac{\mathrm{d}\sigma_T(\gamma h\rightarrow c\bar{c}X
    )}{dz d^2\pv} = \frac{N_c\alpha_{em}e_c^2}{2\pi^2}&\int d^2\mathbf{k}f(x,\mathbf{k})\left\{\left[z^2+(1-z)^2\right]\mathcal{B}_1(\mathbf{p}_T,\mathbf{k})+m_c^2\mathcal{B}_2(\mathbf{p}_T,\mathbf{k})\right\}
\end{align*}
where the auxiliary functions $\mathcal{B}_i$ are defined by
\begin{eqnarray}
    \mathcal{B}_1(\mathbf{p}_T,\mathbf{k})&=&\frac{1}{2}\left[\frac{\mathbf{p}_T}{p^2_T+\epsilon^2}-\frac{\mathbf{p}_T+\mathbf{k}}{(\mathbf{p}_T+\mathbf{k})^2+\epsilon^2}\right]^2 \\
    \mathcal{B}_2(\mathbf{p}_T,\mathbf{k}) &=& \frac{1}{2}\left[\frac{1}{p^2_T+\epsilon^2}-\frac{1}{(\mathbf{p}_T+\mathbf{k})^2+\epsilon^2}\right]^2 \,\,.    
\end{eqnarray}
As a consequence, the charm momentum spectrum is strongly dependent on the modeling of the unintegrated gluon distribution of the nucleus target, which is determined by the QCD dynamics and is sensitive to the description of the nuclear effects. In the next section, we will present our predictions for the UPHIC cross - section derived considering different models for $\mathcal{F}(x,\mathbf{k})$.

\section{Results}
\label{sec:results}
In this section we will present our results for the transverse and rapidity distributions associated with the $D^0$ meson photoproduction in $PbPb$ collisions at  $\sqrt{s} = 5.36$ TeV. In our analysis, we will assume $m_c = 1.4$ GeV and that the fragmentation function is described by the Peterson model \cite{Peterson:1982ak}, which implies  
\begin{equation}
  D^{D^0/c}(z, \mu^2) = \frac{n(D_0)}{z \left[1 - \frac{1}{z} - \frac{\epsilon_c}{1-z}\right]^2}
  \label{placeholder_label} \,\,,
\end{equation}
where $\epsilon_c=0.05$ and {\bf $n(D_0)= 0.308377$}.

It is well known that multiple exchanges of very soft photons between ions can lead to their mutual electromagnetic dissociation. These low-energy photons do not affect the energy--momentum flow in the hard photoproduction process we are interested in
and the respective excitation probabilities factorize. We write the probability that no excitation occurred following\cite{Klusek-Gawenda:2013ema} where more references can be found, as
%%%
\begin{eqnarray}
    P(\rb)  = \exp[ - \bar n (\rb)] \, , \quad  \bar n(\rb) = \int^\infty_{E_{\rm min}}  dE \, N(E,\rb) \, \sigma_{\rm tot} (\gamma Pb; E) \, ,
\end{eqnarray}
%%%
Here, $E_{\rm min}$ is the neutron separation energy, $E$ denotes the photon energy in the nucleus rest frame.
$N(E,\rb)$ should therefore be evaluated with the boost parameter $\gamma_{\rm rf} = 2 \gamma_{\rm cm}^2 -1 $. One then observes that, to a good approximation,
%%%
\begin{eqnarray}
    \frac{E b}{\gamma_{\rm rf}} K_1\Big( \frac{E b}{\gamma_{\rm rf}}\Big) \approx 1,
\end{eqnarray}
%%%
so that
%%%
\begin{eqnarray}
    \bar n(\rb) = \frac{S}{b^2}, \, \quad S =  \frac{Z^2 \alpha_{\rm em}}{\pi^2}\int^\infty_{E_{\rm min}}  \frac{dE}{E} \sigma_{\rm tot} (\gamma Pb; E) \, . 
\end{eqnarray}
%%%%
Using the parameterization of the total photoabsorption cross section of Ref.\cite{Klusek-Gawenda:2013ema}, we obtain for the contribution from the excitation of the giant dipole resonance $S \approx (11.4 \, \rm{fm})^2$, while integrating out to $E = 10 \, {\rm GeV}$, we obtain $S \approx (14.9 \, \rm{fm})^2$.
%{\bf will put my results here...}
%where $\bar n(\rb)$ is given by the total photoabsortption cross section of the lead nucleus
%%%%
%\begin{eqnarray}
%    \bar n(\rb) = \int_{E_{\rm min}}  dE \, N(E,\rb) %\, \sigma_{\rm tot} (\gamma Pb; E) \, ,
%\end{eqnarray}
%%%
%Following Ref.~\cite{Gimeno-Estivill:2025rbw}, we will assume $P(\rb)=\exp(-S/\rb^2)$ and consider two distinct values for the parameter $S$: 
These values lie between  the numbers quoted in \cite{Baur:1998ay}, namely $S=(10.4\;\mathrm{fm})^2$ and $S=(17.4\;\mathrm{fm})^2$, which we adopt in our numerical results below, and which were also used in Ref. \cite{Gimeno-Estivill:2025rbw}. 
For comparison, we will also present the predictions derived assuming $P(\rb)= 1.0$, i.e., summing over all events with and without dissociation.

\begin{figure}[t]
    \centering
    \includegraphics[width=0.49\linewidth]{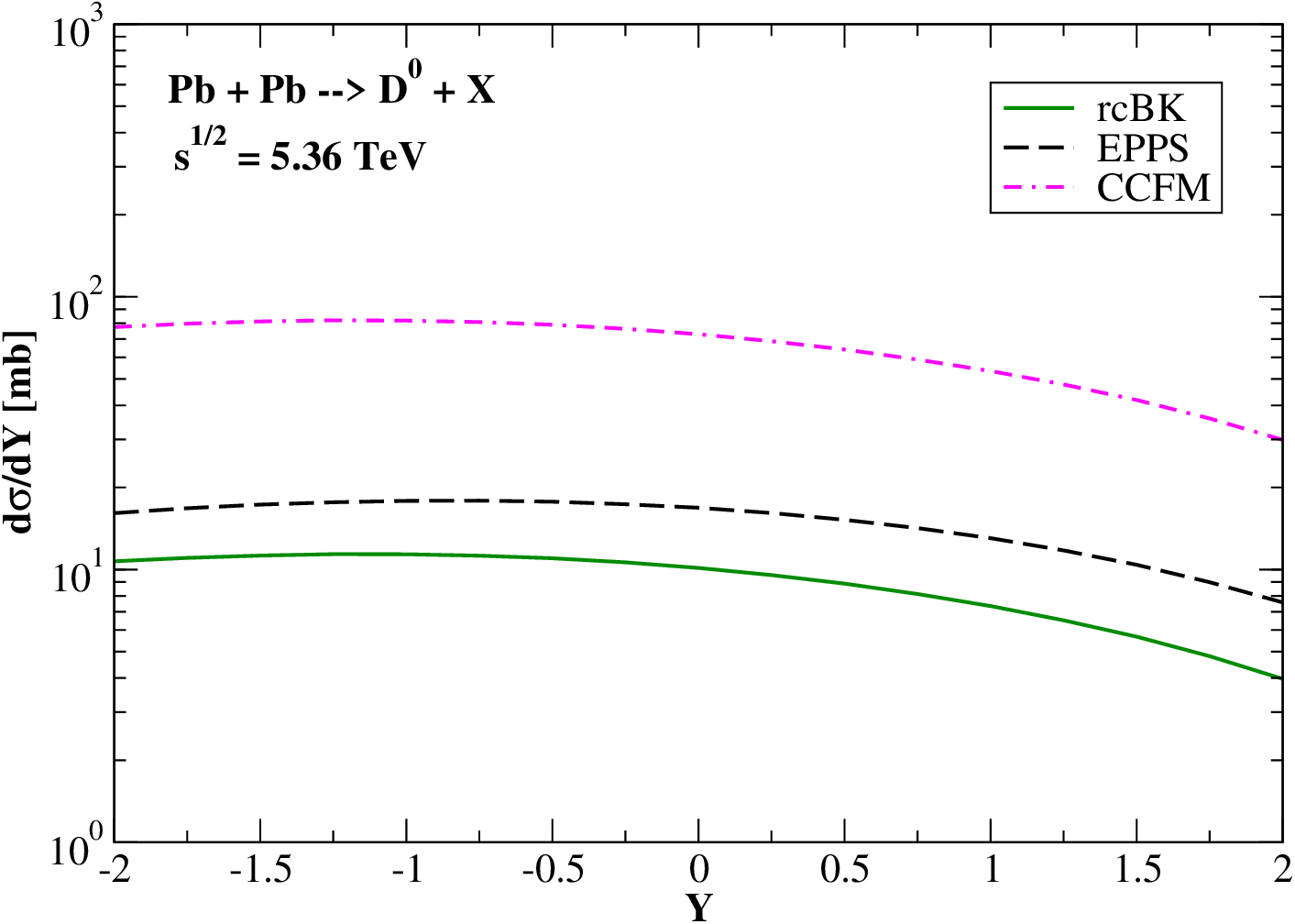}
    \includegraphics[width=0.49\linewidth]{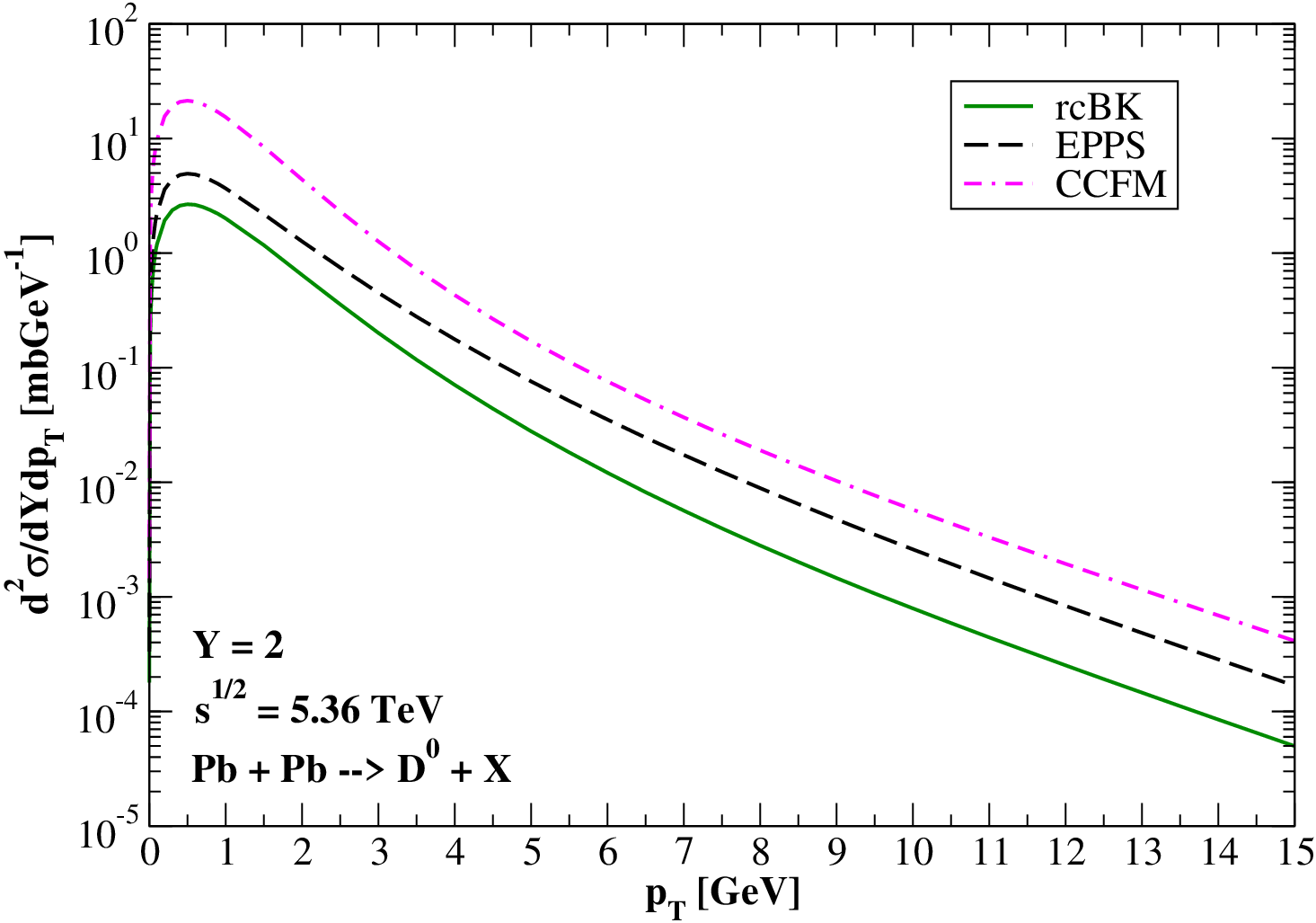}    \caption{Predictions for the rapidity (left panel) and transverse momentum (right panel) distributions associated with the inclusive $D^0$ photoproduction in $PbPb$ collisions at $\sqrt{s} = 5.36$ TeV, derived assuming different models for the nuclear UDG and $P(\rb) = 1$. The nucleus that has emitted the photon was assumed moving in direction of positive rapidities.}
\label{Fig:comparisonUGDs}
\end{figure}

As our goal in this paper is investigate the potentiality of the $D^0$ meson photoproduction in UPHICs to constrain the description of the QCD dynamics and nuclear effects, we will consider three distinct models for $\mathcal{F}(x,\mathbf{k})$, based on very different underlying assumptions. First, we will consider the CCFM-setA1 parametrization for the proton UGD, available in the TMDLIB \cite{Abdulov:2021ivr} \footnote{https://tmdlib.hepforge.org/}, rescaled for a Lead ion, disregarding nuclear effects. Such parametrization was derived in Ref. \cite{Jung:2004gs}  solving numerically the Ciafaloni - Catani - Fiorani - Marchesini (CCFM equation) \cite{Ciafaloni:1987ur,Catani:1989yc,Catani:1989sg} via Monte Carlo method, with the free parameters adjusted in order to describe the HERA data. Such an equation 
 is almost equivalent to BFKL equation in the regime of asymptotic energies and similar to the
DGLAP evolution for large values of $x$ and high values of the hard scale.   Second, we will use the PB-EPPS16 parametrization, also available in the TMDLIB, which was proposed in Ref. \cite{Blanco:2019qbm} considering the parton branching approach and taking into account of the nuclear effects, as predicted by the EPPS16 parametrization \cite{Eskola:2016oht}. These two parametrizations differ in the inclusion (or not) of the nuclear effects, but both are derived assuming that the QCD dynamics is described by a linear evolution equation. Finally, we also will consider the nuclear UGD derived in Refs. \cite{ALbacete:2010ad,Albacete:2012xq}  from the solution of the running coupling Balitsky - Kovchegov (rcBK) equation, which takes into account of nonlinear effects in the   QCD dynamics. 

In Fig. \ref{Fig:comparisonUGDs} we present our predictions for the rapidity and transverse momentum distributions, derived assuming these different models for the nuclear UGD and $P(\rb) = 1$. In this and following figures, we will assume that the nucleus that emits the photon is moving in direction of positive rapidities, which implies that at forward rapidities we are probing small values of $x$ in the nuclear target.  As expected, the CCFM predictions have the larger normalization. The inclusion of the nuclear effects, present in the PB-EPPS16 parametrization, implies the decreasing of the normalization. On the other hand, the rcBK prediction has a smaller normalization with the difference with relation to the PB-EPPS16 one increasing at larger rapidities and transverse momenta.

\begin{figure}[t]
    %\centering
    \includegraphics[width=0.49\linewidth]{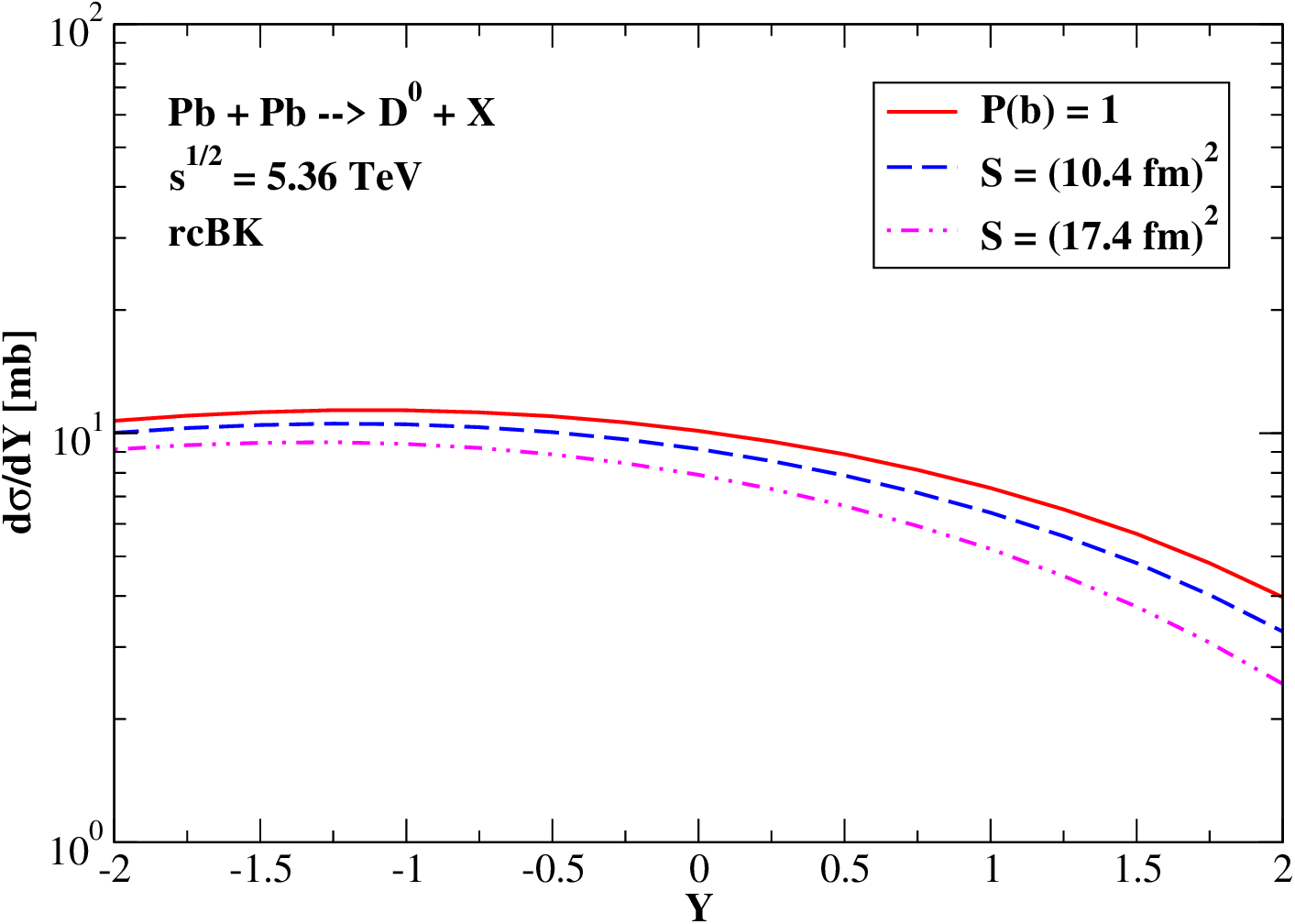}
    \includegraphics[width=0.49\linewidth]{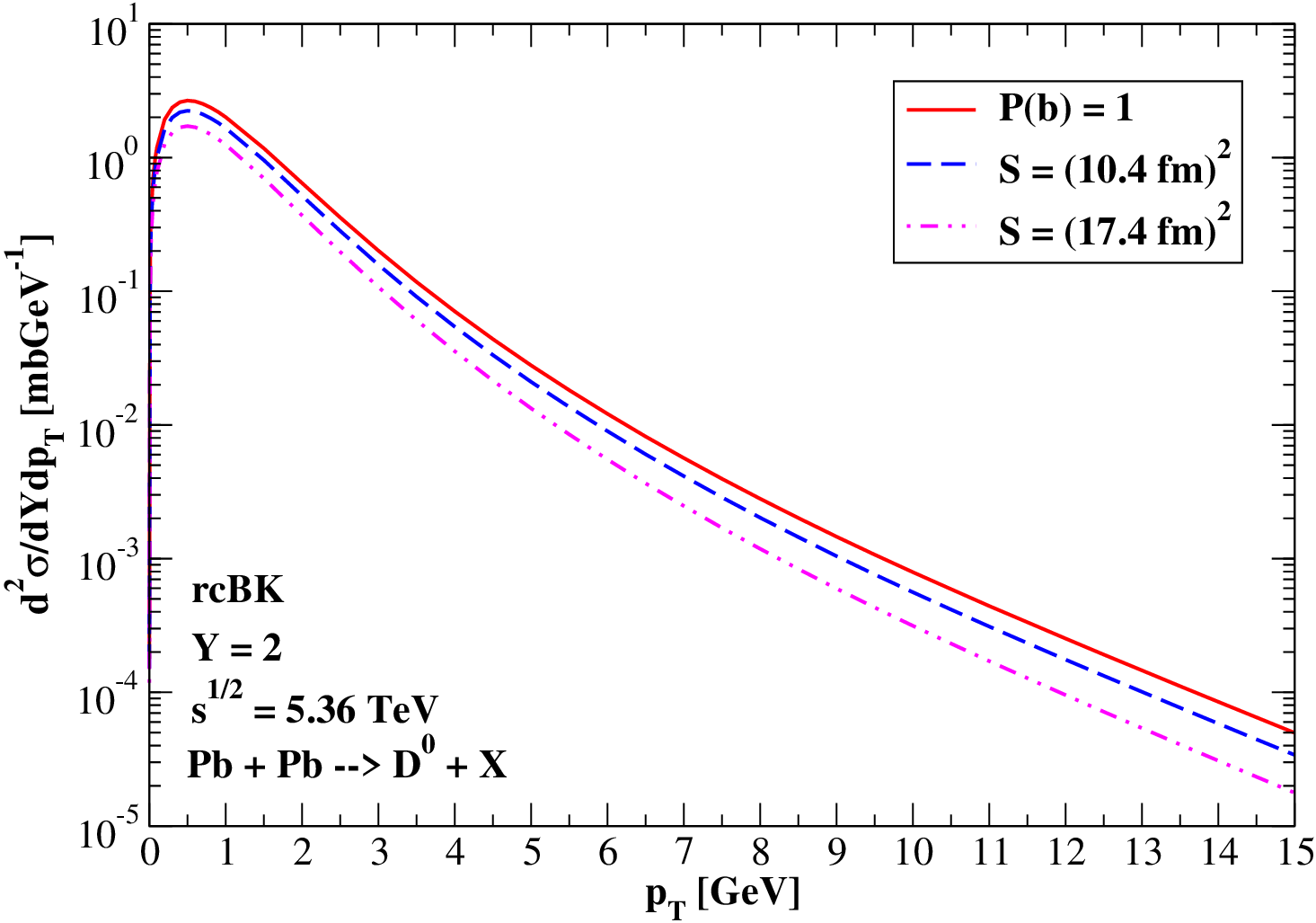}
    \caption{Predictions for the rapidity (left panel) and transverse momentum (right panel) distributions associated with the inclusive $D^0$ photoproduction in $PbPb$ collisions at $\sqrt{s} = 5.36$ TeV, derived assuming the rcBK nuclear UDG and different assumptions for the nuclear electromagnetic dissociation.}
\label{Fig:comparisonAbs}
\end{figure}

In Fig. \ref{Fig:comparisonAbs} we investigate the impact of distinct treatments for $P(\rb)$ on the rapidity and transverse momentum distributions, derived assuming the rcBK nuclear UGD. Similar results were derived for the other two parametrizations, differing only in the normalization. One has that the inclusion of the nuclear dissociation implies a decreasing of the normalization of the distributions, with the impact being larger with the increasing of the parameter $S$. Moreover, our results indicate that the decreasing is larger when the rapidity is increased.

\begin{figure}[t]
    %\centering
    \includegraphics[width=0.32\linewidth]{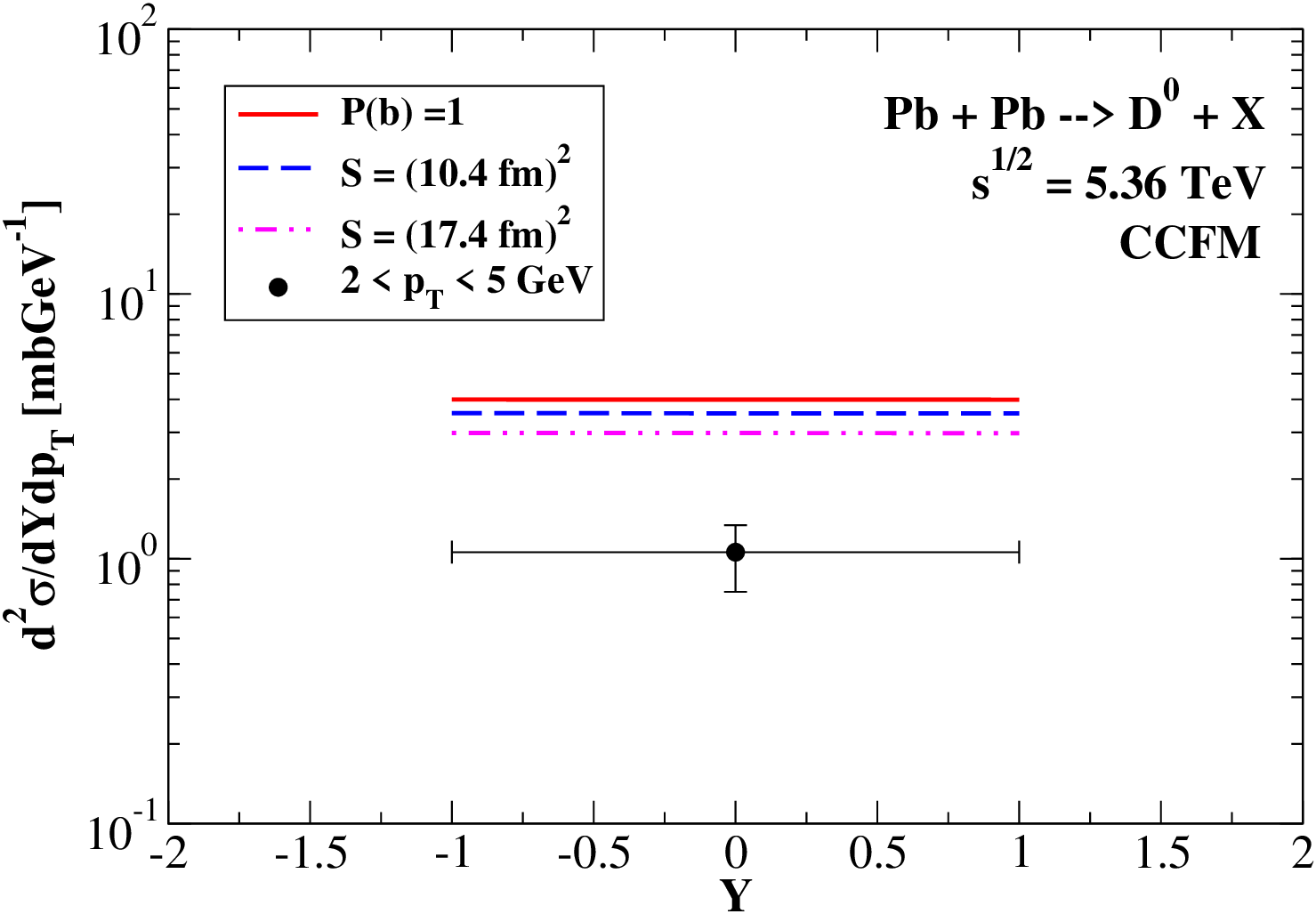}
    \includegraphics[width=0.32\linewidth]{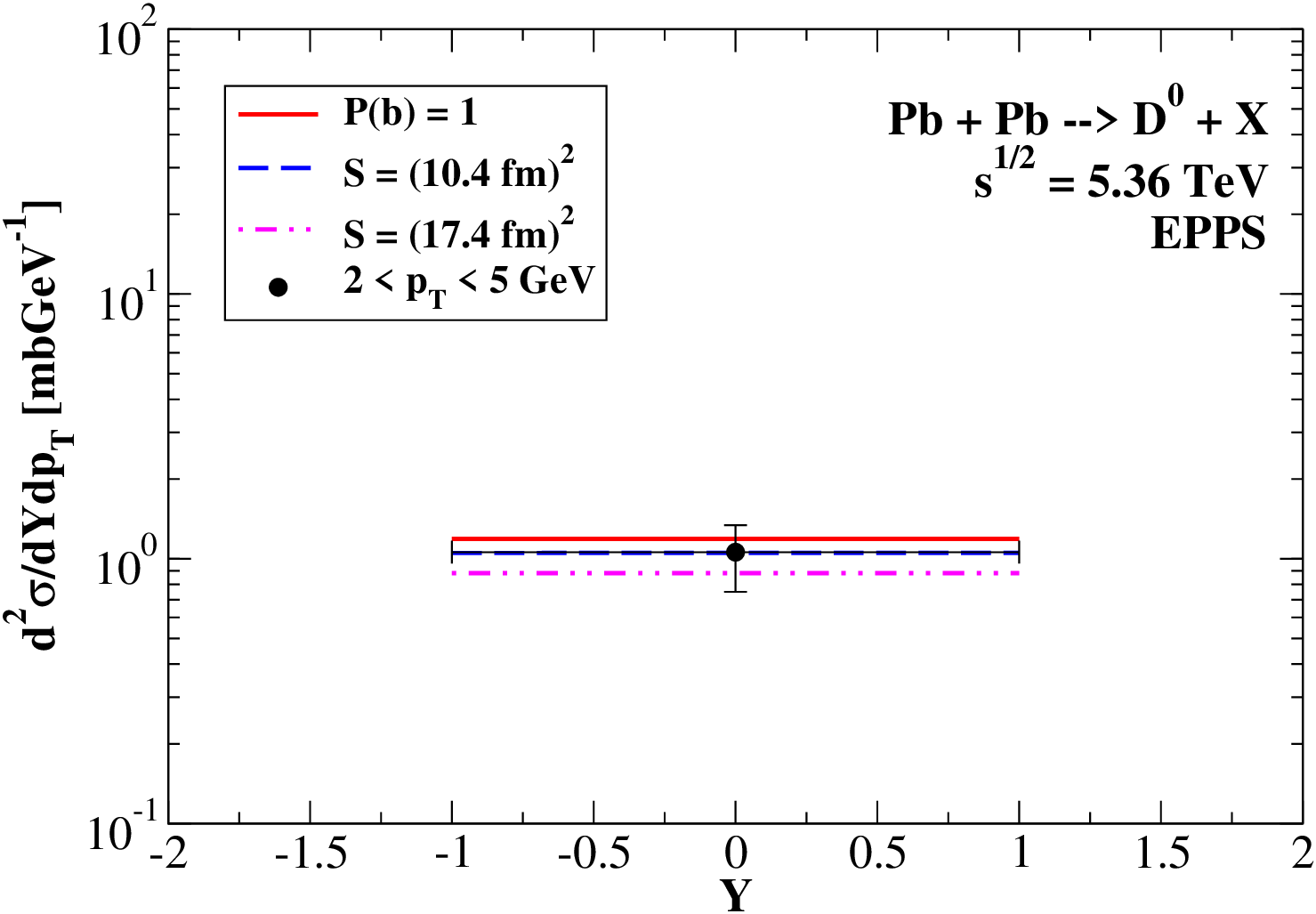}
    \includegraphics[width=0.32\linewidth]{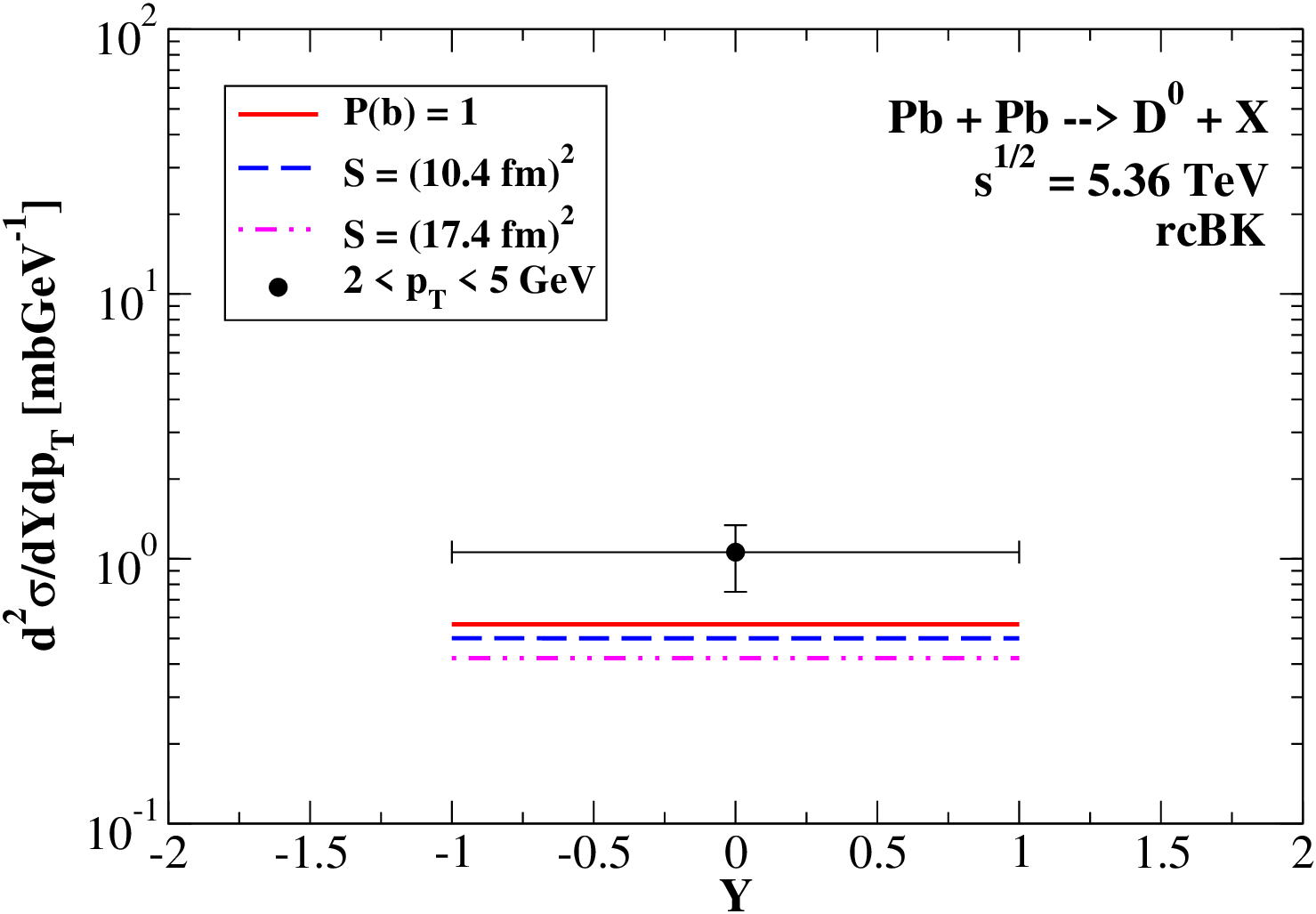} \\
    \vspace{0.8cm}
     \includegraphics[width=0.32\linewidth]{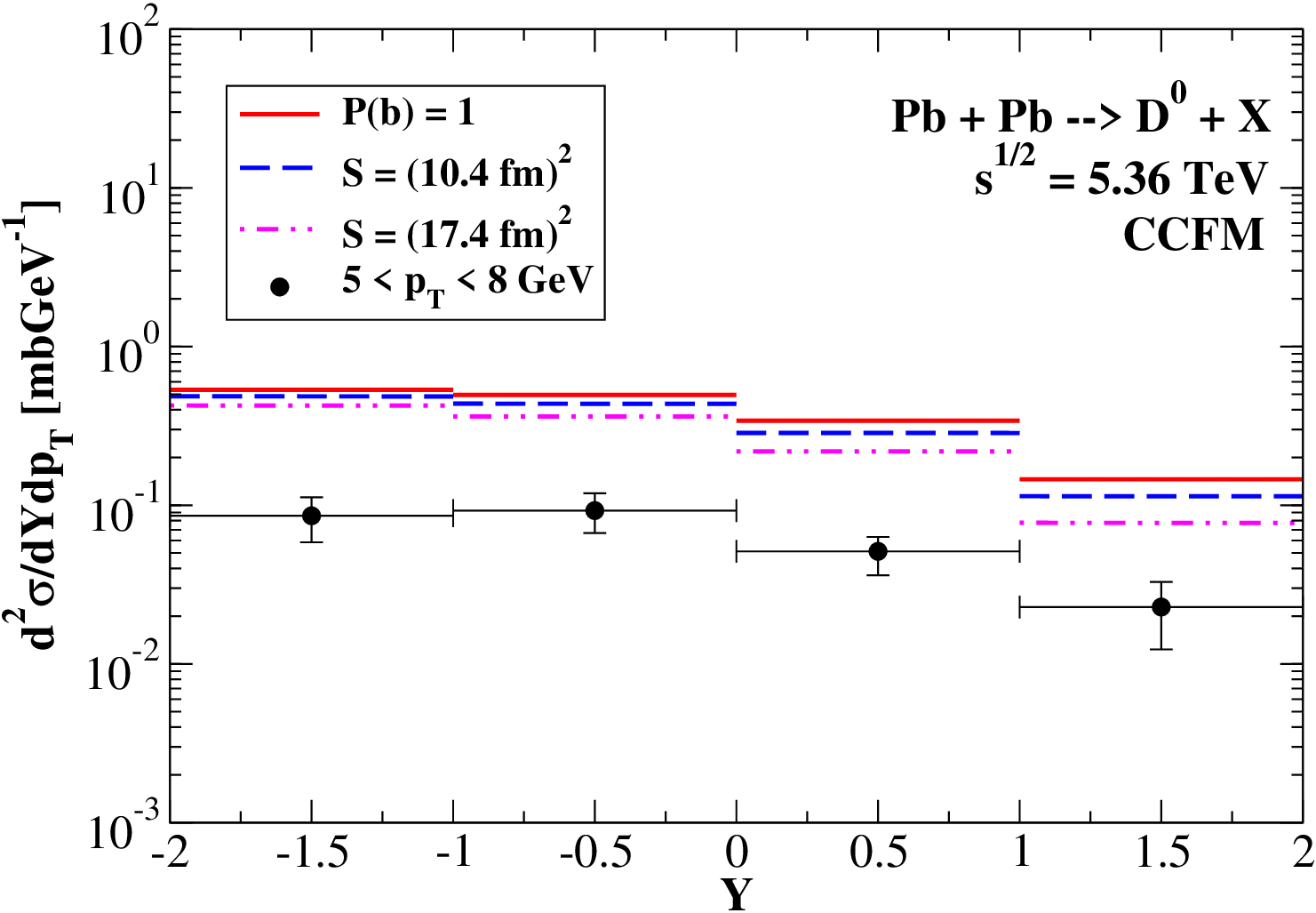}
    \includegraphics[width=0.32\linewidth]{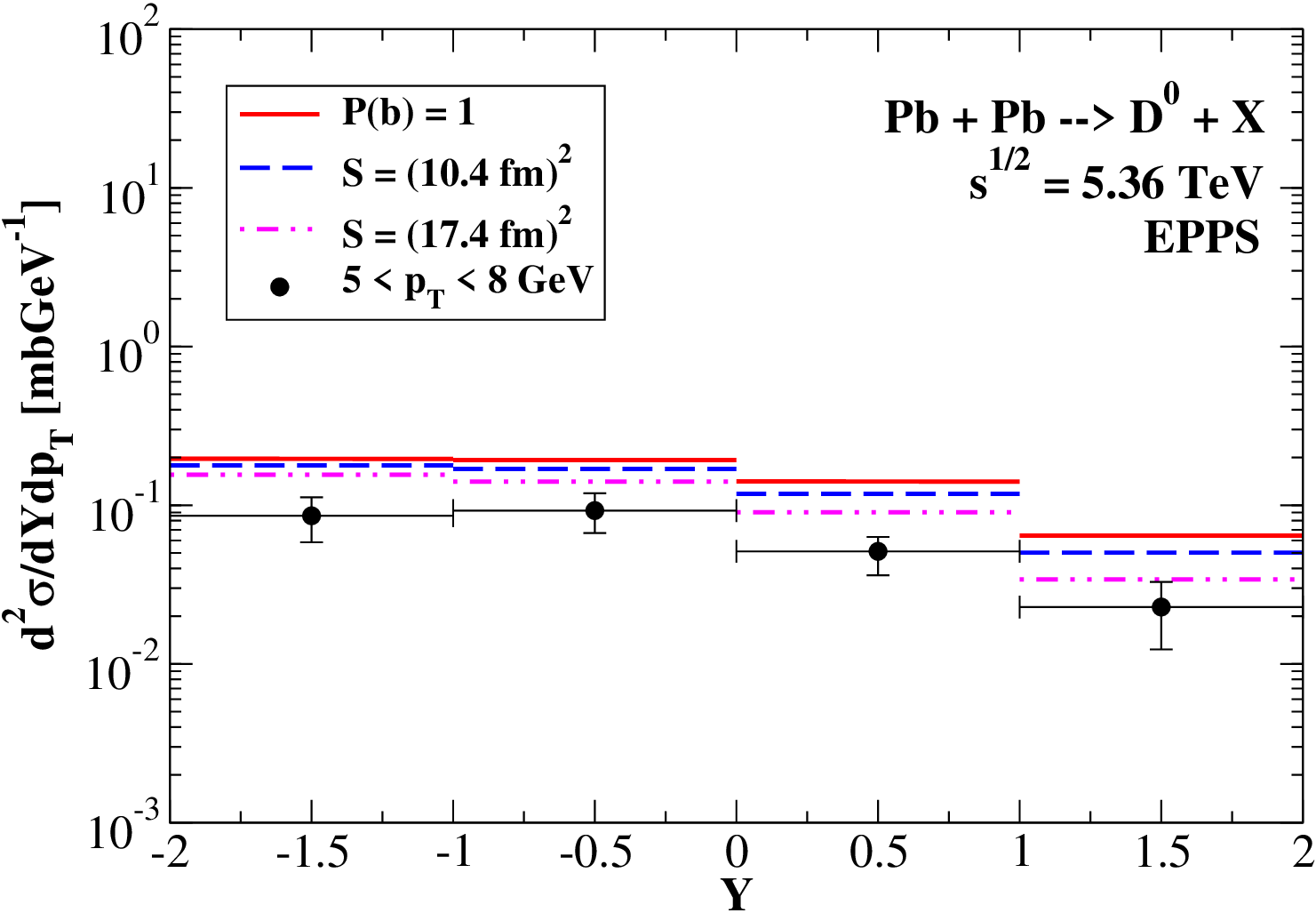}
    \includegraphics[width=0.32\linewidth]{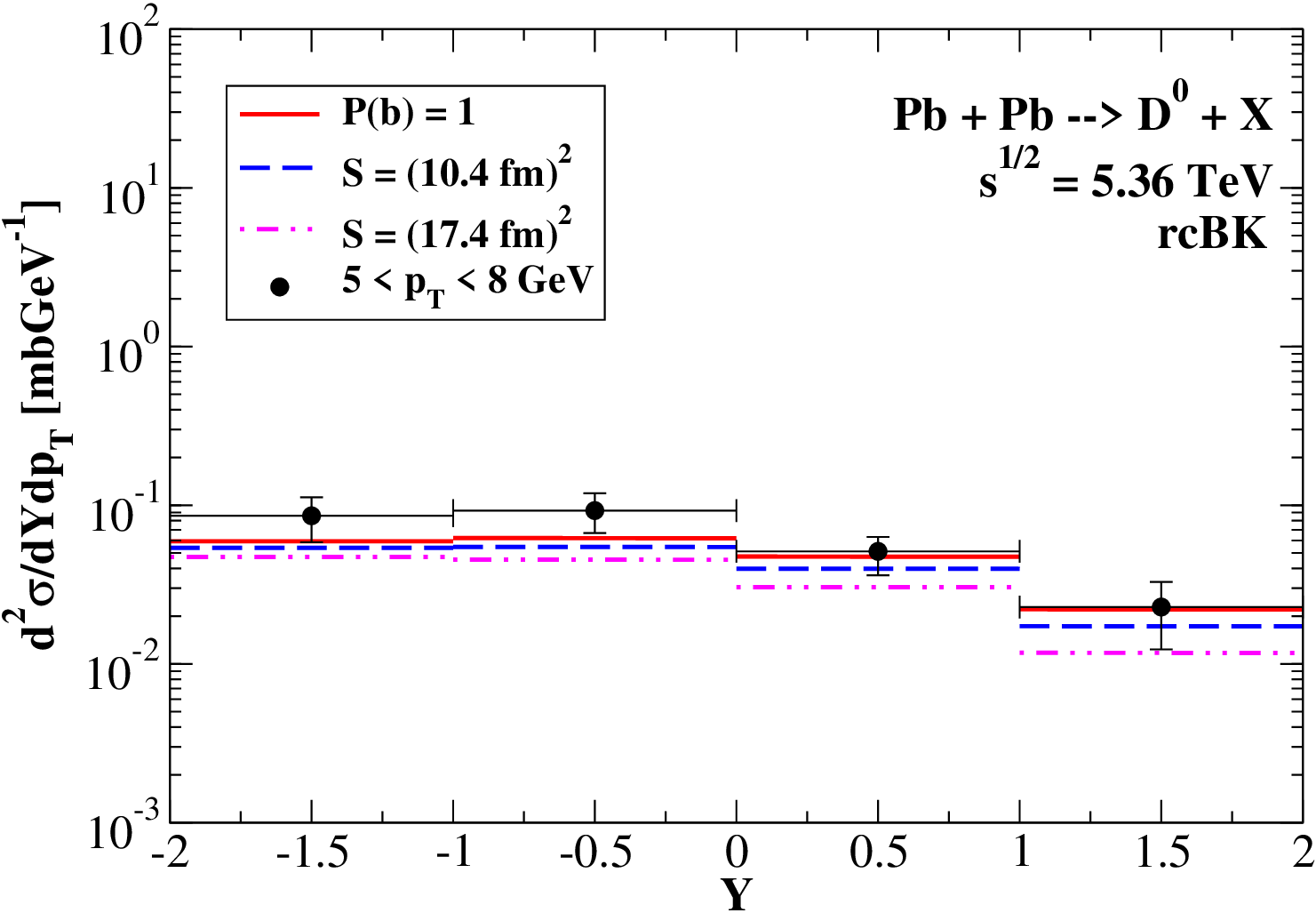} \\
    \vspace{0.8cm}
        \includegraphics[width=0.32\linewidth]{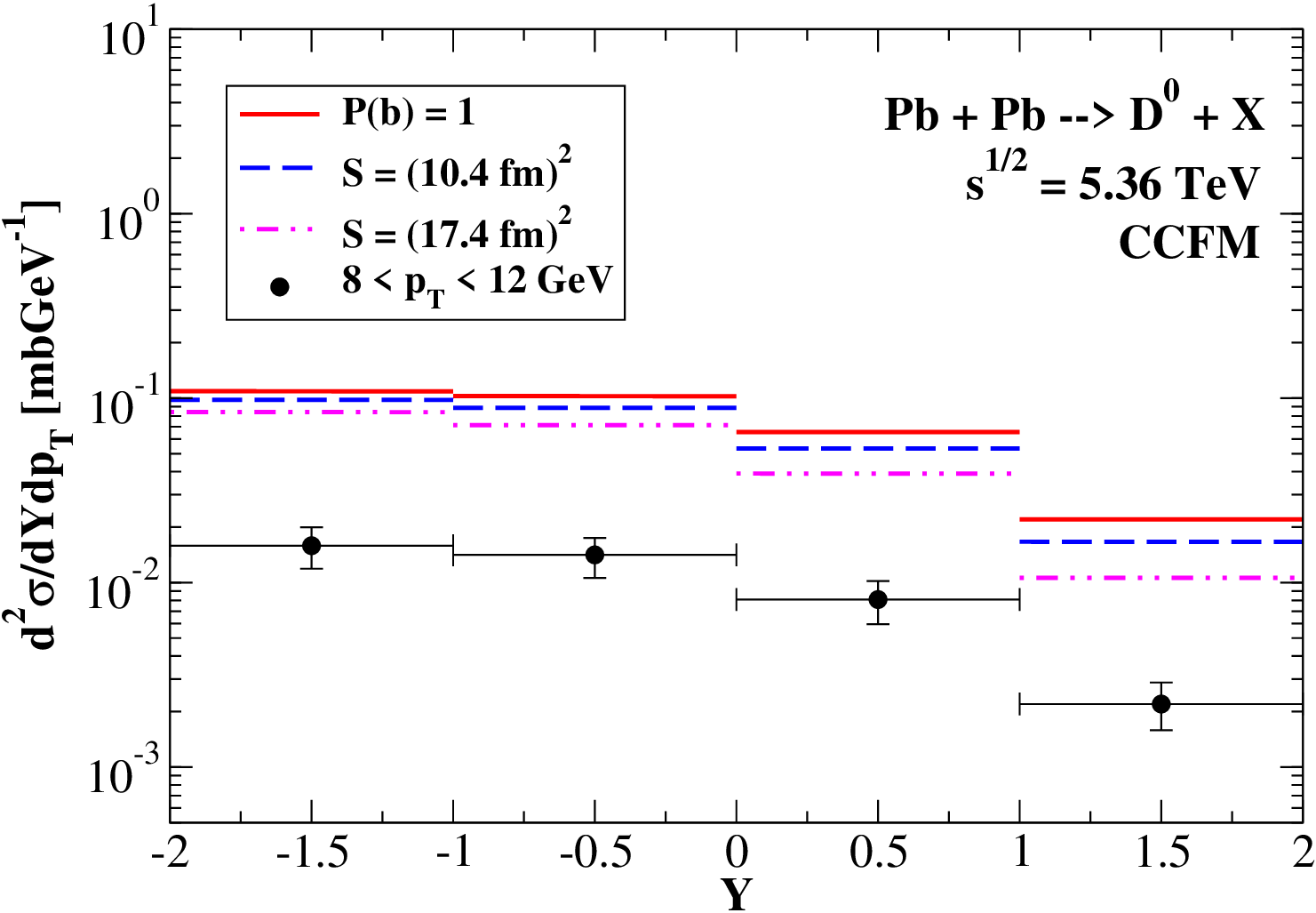}
    \includegraphics[width=0.32\linewidth]{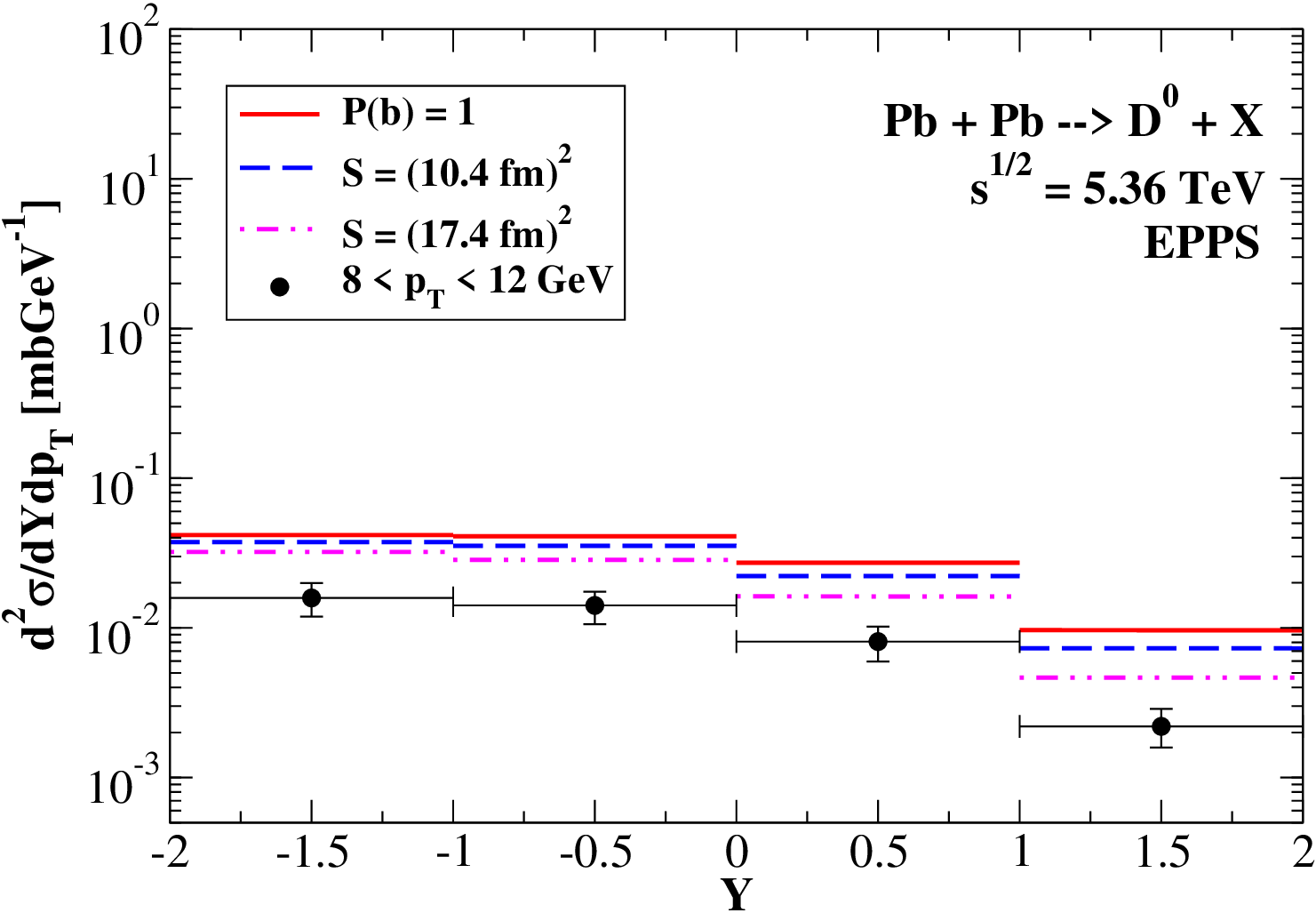}
    \includegraphics[width=0.32\linewidth]{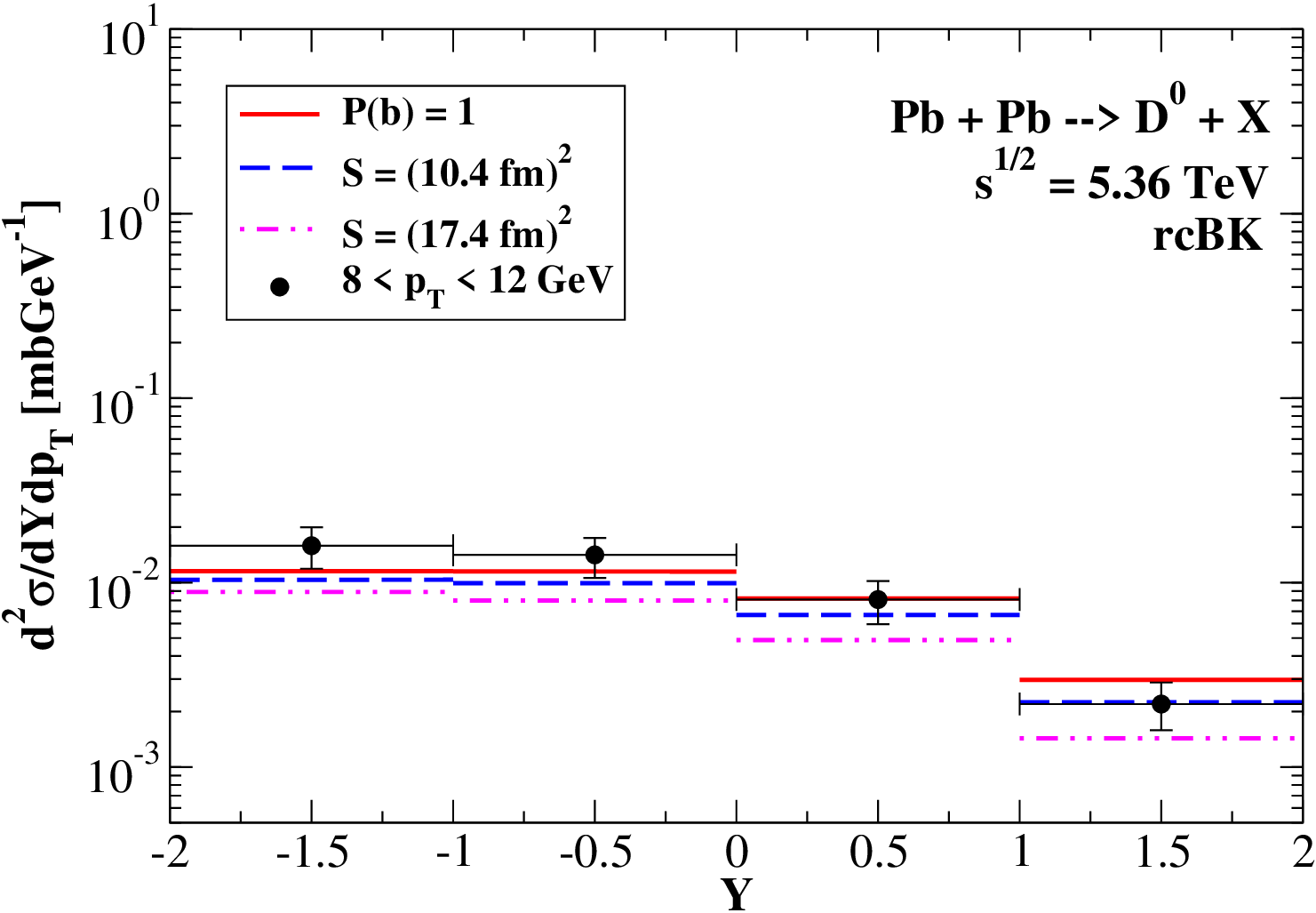}
    \caption{Rapidity distribution associated with the inclusive $D^0$ photoproduction in ultraperipheral $PbPb$ collisions at $\sqrt{s}=5.36$ TeV. Preliminary data from CMS collaboration \cite{CMSpreliminary} obtained considering distint bins in the meson transverse momentum. Theoretical predictions were derived considering distinct models for the nuclear UGD. }
    \label{Fig:CompData}
\end{figure}

In order to compare our results with the preliminary CMS data, we will estimate the average differential distribution, defined by
\begin{eqnarray}
	\left\langle \frac{d^2 \sigma(A + B \rightarrow A \otimes D^0 + X)}{dY_D d^2p_{T,D}} \right\rangle = \frac{1}{\Delta p_T} \frac{1}{\Delta y} \int_{y_{min}}^{y_{max}} dy  \int_{p_T^{min}}^{p_{T}^{max}}  d p_{T,D} \frac{d^2 \sigma(A + B \rightarrow A \otimes D^0 + X)}{dY_D d^2p_{T,D}}  \,\,,
\end{eqnarray}
with $\Delta p_T = p_{T}^{max} - p_T^{min}$ and $\Delta y = y_{max} - y_{min}$, where $p_T^{max}$ ($y_{max}$)  and $p_T^{min}$ ($y_{min}$) are the upper and lower values of the transverse momentum (rapidity) in a given bin. In Fig. \ref{Fig:CompData}, we present a comparison of our predictions with the recent (preliminary) CMS data \cite{CMSpreliminary} for the rapidity distribution considering different transverse momentum bins associated with the inclusive $D^0$ photoproduction in ultraperipheral $PbPb$ collisions at $\sqrt{s} = 5.36$ TeV. Our results have been derived considering the different models for the nuclear UGDs. One has that  the CCFM prediction, which disregard nuclear effects, is not able to describe the data. In contrast, the other two models provide a better description, which indicate that  nuclear effects and/or nonlinear effects in the QCD dynamics must be included to describe the data. Surely, more precise data, derived considering smaller bins in $p_T$ and a larger range of rapidity would be very useful to improve our description of the nucleus at small - $x$.

\section{Summary}
\label{sec:summary}
The description of the QCD dynamics at high energies is still a theoretical challenge. Several models are available in the literature that consider distinct approaches and different underlying assumptions to treat the dynamics of a high gluon density system and how it is modified by the presence of a nuclear medium. In order to advance in our understanding, it is fundamental to have data for observables that are sensitity to these aspects.
In this paper, we have performed an exploratory study of the inclusive $D^0$ photoproduction in UPHICs. We have derived the corresponding differential cross - section considering the color dipole $S$ - matrix approach, and demonstrated that it is determined by the unintegrated gluon distribution of the nuclear target. Predictions for the rapidity and transverse momentum distributions were derived considering three distinct models for the UGD and the results indicate that these observables are very sensitive to the descriptions of the QCD dynamics and nuclear effects. Moreover, a comparison with recent (preliminary) CMS data has been performed. Our results indicate that precise experimental data for this observable will be very useful to improve our understanding of the strong interaction theory. 

\begin{acknowledgments}
V.P.G. is grateful to the Mainz Institute of Theoretical Physics (MITP) of the Cluster of Excellence PRISMA+ (Project ID 390831469), for its hospitality and  support.  V.P.G. and L. S.  were partially supported by CNPq, CAPES, FAPERGS and INCT-FNA (Process No. 464898/2014-5). 
The work of W.S. was partially supported by 
the Polish National
Science Center Grant No. UMO-2023/49/B/ST2/03665.
\end{acknowledgments}

\bibliographystyle{unsrt}

\end{document}